\documentclass[aps,prl,twocolumn,groupedaddress]{revtex4-2}

\usepackage{amsmath,amssymb}
\usepackage{graphicx}
\usepackage{bbold}
\usepackage[colorlinks=true, allcolors=blue]{hyperref}
\usepackage[dvipsnames]{xcolor}
\def\H{\mathcal{H}}
\def\k{\boldsymbol{k}}

\def\q{{\bf q}}
\def\d{^{\dagger}}
\def\up{\uparrow}
\def\down{\downarrow}
\def\D{\Delta}
\def\cc{^{\ast}}

\def\s{\sigma}

\begin{document}


\title{Altermagnetism in an interacting model of Kagome materials}


\author{Alejandro Blanco Peces}
\email[]{alejandro.blancop@uam.es}
\author{Jaime Merino}
\email[]{jaime.merino@uam.es}
\affiliation{Departamento de Física Teórica de la Materia Condensada, Condensed Matter Physics Center, (IFIMAC) and Instituto Nicolás Cabrera, Universidad Autónoma de Madrid, 28049 Madrid, Spain}


\date{\today}

\begin{abstract}

The Hubbard model on the Kagome lattice is a widely used interacting model for describing the electronic properties of various transition metal-based Kagome materials. We find altermagnetism driven by Coulomb interaction in the Kagome Hubbard model at Dirac filling with no spin-orbit coupling nor explicit spatial symmetry breaking present. We show how this insulating altermagnet is relevant to other lattices with larger unit cells such as the Lieb-Kagome lattice. The ALM found displays a characteristic magnon splitting which can be detected in inelastic neutron scattering experiments on interacting Kagome materials.

\end{abstract}


\maketitle

\emph{Introduction.}
Quasi-two-dimensional transition metal Kagome 
materials display a rich variety of electronic phases 
due to an interplay between electronic correlations,  geometric frustration and topology. While AV$_3$Sb$_5$ (A=K, Rb, Cs) \cite{Ortiz2019,Nguyen_2022} displays $2 \times 2$ charge density wave (CDW) order, superconductivity \cite{kang_charge_2023,hu_electronic_2023, Jiang_2022, Ortiz2020, Mielke_2022, Li2021} and a giant anomalous Hall effect \cite{Yang_giant_2020,jiang_unconventional_2021, Yu2021}, these materials are considered weakly correlated and close to van Hove filling, while the Dirac filled 
ScV$_6$Sn$_6$ displays $\sqrt{3}\times \sqrt{3}$ CDW order\cite{hu_phonon_2024,Arachchige2022,Lee2024_2}.
In strongly correlated Kagome materials hosting flat bands a spin density wave (SDW) typically accompanies the CDW order as in AFM FeGe displaying $2 \times 2$ CDW order\cite{Dai_nature_2022,Teng_nature_physics_2023} and in 
CsCr$_3$Sb$_5$\cite{Li_nature_commun_2025,liu_superconductivity_2024} where, under an external pressure, the spin/charge DW can be suppressed giving way to superconductivity.
Under moderate pressures or small hole dopings, the CDW in CsV$_3$Sb$_5$ is suppressed and two possibly unconventional \cite{hossain_unconventional_2025, yu_natcom_2021} superconducting domes emerge. The intertwinning between CDW/SDW, and superconductivity observed in Kagome materials is reminiscent of high-T$_c$ superconductors raising similar questions about the mechanism of superconductivity and the nature of the unconventional metallic state. It is interesting to explore whether other forms of collinear magnetism different from FM or AFM, such as altermagnetism, can emerge in these materials. 

Altermagnetism, a novel form of magnetic order with an underlying collinear antiferromagnetic order breaking $\mathcal{PT}$-symmetry has been proposed to occur in several materials \cite{Hayami_momentum_2019,naka_spin_2019,sinova_crystal_2020,mazin_prediction_2021, cheong_npj_2024,Lee2024}. Apart from their fundamental importance they can generate spontaneous spin currents even without spin-orbit coupling so they can be relevant to spintronics. In contrast to conventional AFMs, sites with opposite spins in an altermagnet (ALM) are not related by inversion or lattice translation but by lattice rotation and/or reflection symmetries. This leads to the breaking of $\mathcal{PT}$-symmetry and consequently to a momentum-dependent spin-splitting of the bands \cite{Mazin2022,Smejkal2022,Smejkal2022_2,Krempasky2024}. Strikingly, the DW observed in CsCr$_3$Sb$_5$ \cite{xu_altermagnetic_2025} has been found to be ALM based on \emph{ab initio} calculations. This raises fundamental questions about the origin of altermagnetism in Kagome materials, its possible connection with the nearby superconducting state induced by pressure and its relevance to the close-by non Fermi liquid state \cite{liu_superconductivity_2024}.

\begin{figure}
\includegraphics[width=\columnwidth]{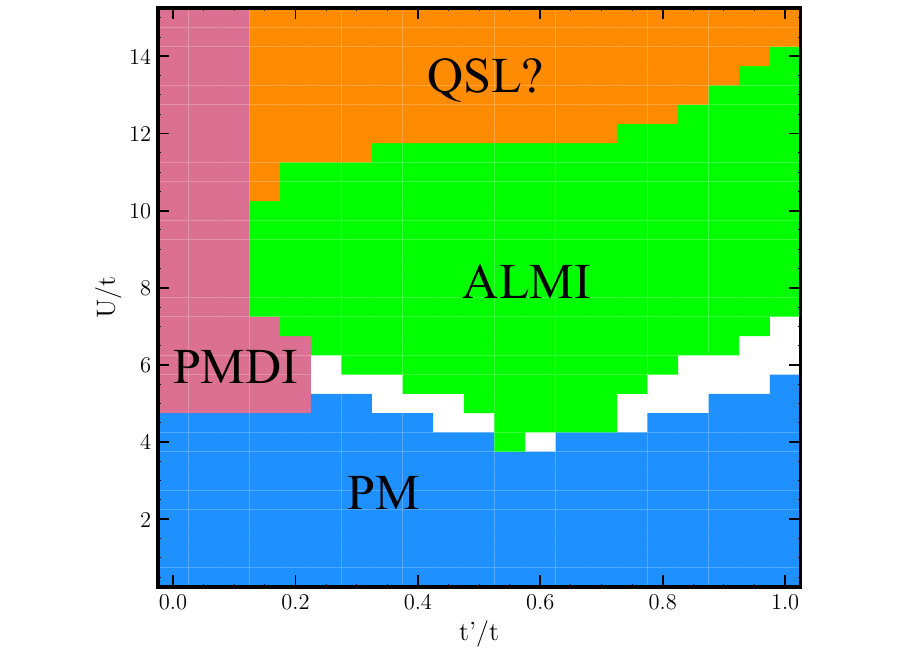}
\caption{\label{Ut2} Altermagnetism in the Kagome Hubbard model at Dirac filling. The $t'-U$ phase diagram at $n=2/3$ and fixed temperature $T=0.02t$ obtained from HF is shown.  
A paramagnetic metal (PM), an altermagnetic insulator (ALMI), a pinned metal droplet insulating (PMDI), and a possible quantum spin liquid (QSL) phase arise. In the white regions stable converged phases are not found. HF calculations on $N\times N$ cell lattices with $N=12 - 18$ and periodic boundary conditions have been used.}
\label{fig:pd}
\end{figure}

\begin{figure*}
\includegraphics[width=\linewidth]{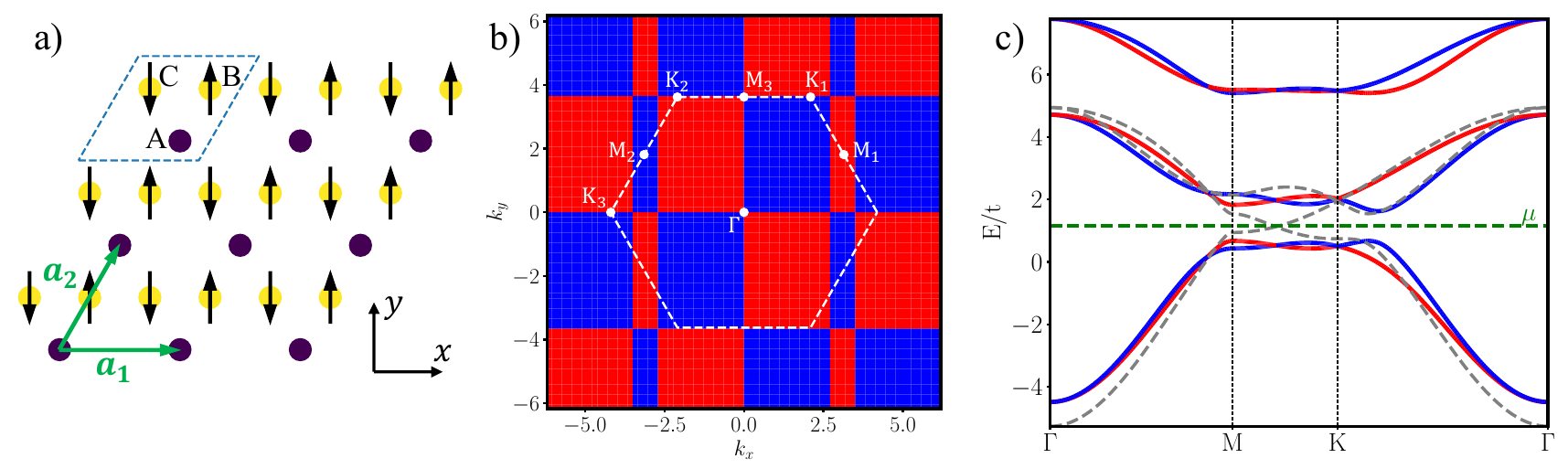}
\caption{ a) Charge and spin densities in the ALM state of the Kagome Hubbard model at $n=2/3$, $U=8$ and $t'=0.7t$. Yellow (purple) sites indicate higher (lower) charge densities and arrows show mean spin vector directions; the unit cell is enclosed by blue dashed lines. b) Spin character of  energy eigenvalues in the topmost band. Red and blue regions correspond to spin up and down eigenstates, respectively; the dashed white lines mark the edges of the first Brillouin Zone. c) Energy bands of the Kagome ALM along a high-symmetry path in momentum space. Again, red and blue lines indicate the spin up and down character of the bands at the corresponding momentum. The chemical potential is shown with a dashed green line, and dashed gray lines show the (spin degenerate) tight-binding band structure of the Kagome lattice.}
\label{fig:Kagome_altermagnet}
\end{figure*}

In this Letter, we establish the existence of altermagnetism arising spontaneously in a spatially uniform single-band Kagome and Lieb Hubbard models. The breaking of the $\mathcal{PT}$-symmetry occurs spontaneously in a uniform Hubbard model with just one orbital per site. 
The ALM found has ${\bf q}=0$ order and is stable in a broad parameter (see Fig. \ref{fig:pd}) and temperature range. We show how the insulating ALM reported is not specific of the Kagome lattice but is generally present in Lieb and Lieb-Kagome lattices consisting on larger unit cells with an odd number of sites $N_c$ hosting an even number of electrons. Under this condition, if $\mathcal{PT}$-symmetry is broken through AFM order, a gap is opened stabilizing the ALMI. Unlike in conventional AFMs, magnon excitation energies in the ALMI are splitted according to their chirality.
Our finding differs from recent proposals which consider Hubbard models that explicitly break translational symmetry\cite{franz_prl_2025,thomale_prl_2025,Sudbo_2023}, rotational sublattice symmetry\cite{Maier2023,Antonenko2025}, or include several orbitals per site\cite{Capone_altermagnetism_2025}, longer range Coulomb interaction \cite{Wang_altermagnetism_2025} or 
orbital altermagnetism generated by loop currents \cite{Fernandes2025}.
Although ScV$_6$Sn$_6$ is an $n=2/3$ Kagome material \cite{hu_phonon_2024}, it hosts a $\sqrt{3} \times \sqrt{3}$ CDW resembling the PMDI of Fig. \ref{fig:pd} but with no magnetic order. Thus, Cr-based Kagome materials with $n=2/3$ and appreciable $t'$ are candidates for the ALMI found here. 

\emph{Model and phase diagram.}
The simplest model to understand the electronic properties of Kagome materials is the Kagome Hubbard model:
\begin{equation}
    \H=\sum_{ij}t_{ij}c\d_{i\sigma}c_{j\sigma}+U\sum_j n_{j\up}n_{j\down},
    \label{eq:hubbard}
\end{equation}
where $c\d_{i\sigma}$ ($c_{i\sigma}$) creates (annihilates) an electron at site $i$ with spin $\sigma=\{\up,\down\}$, and $t_{ij}$ is the hopping amplitude between sites with $t_{ij}=-t$ when $ij$ are nearest-neighbors (n.n.) and $t_{ij}=-t'$ when they are next-nearest-neighbors (n.n.n.). $U$ 
is the onsite Hubbard repulsion and $n_{i\sigma}=c\d_{i\sigma}c_{i\sigma}$ is the density operator. Model (\ref{eq:hubbard}) differs from Hubbard models on modified Lieb lattices containing an extra onsite potential naturally leading to CDW order\cite{franz_prl_2025,thomale_prl_2025, Sudbo_2023}. We perform an unrestricted Hartree-Fock (HF) treatment of model (\ref{eq:hubbard}) on large finite lattices \cite{Metzner2023, Moore2024} complemented with solutions valid in the thermodynamic limit (see Supplementary Material). At fixed $U$ and electronic density, $n$, we can obtain the real space charge and spin patterns for any temperature, $T$. 

In Figure \ref{Ut2} we show the $t'-U$ phase diagram of model (\ref{eq:hubbard}) at $n=2/3$ and at low temperatures, $T=0.02t$. At weak coupling, $U \lesssim 4t$, the system is a paramagnetic metal (PM) for any $t'/t$. As $U$ is increased a metal-insulator transition occurs. For low $t'/t$ a pinned metal droplet insulator (PMDI) which is stable up to large $U$ arises \cite{Ralko2014, Kim_2020}. For larger $t'/t \gtrsim 0.15$ an altermagnetic insulator (ALMI) emerges (see Fig. \ref{fig:Kagome_altermagnet} (a)) for non-zero $t'/t$ which covers a broad $U$ region of the phase diagram. At strong coupling ($U\gtrsim 12t$), the ALMI is unstable eventually becoming a disordered quantum spin liquid (QSL) consistent with exact diagonalization (ED) results \cite{McCullloch2015} (see Supplementary Material).  



The ALMI found is a $\q=\bf 0$ state with intracell charge and spin disproportionation (see Fig. \ref{fig:Kagome_altermagnet}a)). One site in the unit cell has low charge density and negligible spin whereas the other two have equal larger charge densities and opposite spins. This leads to a zero magnetization state as expected in an ALM. In addition, to classify as an ALM, the real space configuration in Fig. \ref{fig:Kagome_altermagnet}a) should break $\mathcal{PT}$ symmetry so we turn our attention to the symmetries in the ALMI.  While the C$_6$ rotation symmetry around the center of the hexagon is reduced to a two-fold C$_2$ symmetry, the $C_3$ symmetries around the two inequivalent triangles are completely broken while the $C_2$ symmetry around each lattice site is preserved. Hence, the ALMI is not invariant under $\mathcal{T}$ inversion leading to spin inversion and spatial $\mathcal{P}$ inversion or translation as in a conventional AFM. 


Our phase diagram also highlights the prominent role played by a nonzero $t'$ \cite{Wang2025} in stabilizing our ALMI phase. Indeed, at vanishing $t'$, the PMDI, a collinear spin
configuration consisting on a combination of a $\sqrt{3}\times\sqrt{3}$ CDW and a SDW, not breaking $C_6$ symmetry, wins. Although the PMDI has zero net magnetization, it is not ALM since it does not break $\mathcal{PT}$ symmetry.

\emph{Electronic structure of altermagnetic phase.}
The momentum dependent spin splitting of the bands characteristic of ALMs is evident in Fig. \ref{fig:Kagome_altermagnet}c) where various colors indicate the spin character of the energy eigenstates. The presence of momentum-dependent spin-split bands confirms that the system is ALM. Spontaneous breaking of the $C_6$ translates into the specific spin-splitting pattern (see Fig. \ref{fig:Kagome_altermagnet}b)) consisting of alternating spin-up and spin-down patches with only C$_2$ symmetry around the $\Gamma$ point (see Fig. \ref{fig:Kagome_altermagnet}b)) consistent with the real space pattern. In addition, the band structure shows that the ALM is an insulator, in contrast to the original non-interacting semimetal with two Dirac cones inside the Brillouin Zone. 

The origin of the spin-splitting can be better understood with a minimal model of the Kagome lattice with a site-dependent Zeeman field $\boldsymbol{J}_i$:
\begin{equation}
    \H=\sum_{ij}t_{ij}c\d_{i\s}c_{j\s}+\sum_i \boldsymbol{J}_i\cdot\boldsymbol{S}_i
\end{equation}
where $\boldsymbol{S}_i$ is the spin operator on site $i$.  We could also include a site-dependent chemical potential, $\mu_i$, to reproduce the charge imbalance between the spin-zero and the non-zero spin sites and fully retrieve the ALM charge pattern.
Although any site-dependent potential: $\mu_A \neq \mu_B=\mu_C$ would automatically open a gap at the Fermi level, the $\boldsymbol{J}_i$ are enough to explain the key features characterizing the electronic spectrum of the ALM: the gap and the spin-splitting. As the ALM consists of collinear spins pointing along an arbitrary axis, it is equivalent (related by a rotation) to a state where the nonzero spins point along the $\pm z$ directions, so that 
the spin-up and spin-down sectors are uncoupled:
\begin{equation}
    \H=\sum_{i,\sigma,\k}c\d_{i\sigma\k}(H_{\sigma\k})_{ij}c_{j\sigma\k}
\end{equation}
where $c^{(\dagger)}_{i\sigma\k}$ is the Fourier Transform of $c^{(\dagger)}_{i\sigma}$ and we have defined the matrix
\begin{equation}
    H_{\sigma\k}=\begin{pmatrix}
        0 & \epsilon_{AB} & \epsilon_{AC} \\
        \epsilon_{AB}\cc & J\s & \epsilon_{BC} \\
        \epsilon_{AC}\cc & \epsilon_{BC}\cc & -J\s
    \end{pmatrix}
\end{equation}
with $\s=\{\up,\down\}\equiv\{+1,-1\}$, where $\epsilon_{ij}$ are the momentum-dependent hopping amplitudes between sites $ij$ (we have dropped the $\k$-dependence for simplicity).
Within this simple model, the spin is zero on $A$ and opposite on $B$ and $C$. The energy eigenstates $E_{\s\k}$ are  given by the solutions of the eigenvalue equation $\det(H_{\s\k}-E_{\s\k}\mathbb{I})=0$ (dropping $\k$ subscripts):
\begin{multline}
    -E_{\s}^3+(|\epsilon_{AB}|^2+|\epsilon_{AC}|^2+|\epsilon_{BC}|^2+J^2)E_{\s}\\ +2\Re(\epsilon_{AC}\epsilon_{AB}\cc\epsilon_{BC}\cc)+J\s(|\epsilon_{AB}|^2-|\epsilon_{AC}|^2)=0
    \label{eq:eigen}
\end{multline}

In general, the eigenvalue equations for the spin-up ($\s=+1$) spin-down ($\s=-1$) sectors are different due to the last term in the Eq. (\ref{eq:eigen}) so there will be a spin-splitting at those momenta at which $|\epsilon_{AB}| \neq |\epsilon_{AC}|$.

Although $|\epsilon_{AB}|$ and $|\epsilon_{AC}|$ have the same functional form on the Kagome lattice due to the original equivalence of the sites, they are related by a rotation around $\Gamma$ and so they are not equal except at a set of lines in momentum where the spin-splitting vanishes and a change in the spin character of the eigenvalues occurs. The above minimal model also shows that $J$ opens a gap at any band crossing of the original non-interacting bands, as shown in Fig. \ref{fig:Kagome_altermagnet}c). Due to the 
even filling of the ALMI, we conclude that it is an insulator (see supplementary for details). 

It should be noted that this type of ALM cannot occur on a two-site unit cell lattice, where ALM can be realized at the cost of  breaking the equivalence between the two sites which can be achieved by introducing two different intra-sublattice hoppings ($|\epsilon_{AA}|\neq|\epsilon_{BB}|$). This would lead to terms linear in $J$ in the eigenvalue equation. Thus, our $U$-driven ALMI is qualitatively different from previous proposals limited to two-site unit cells \cite{Maier2023,Roig2024}. In brief, the type of ALM described here requires an odd number of sites in the unit cell and an even number of electrons, in a configuration such that opposite spin sites are not related to each other by inversion w.r.t. a zero-spin site, as further discussed below.


\begin{figure}
\includegraphics[width=\columnwidth]{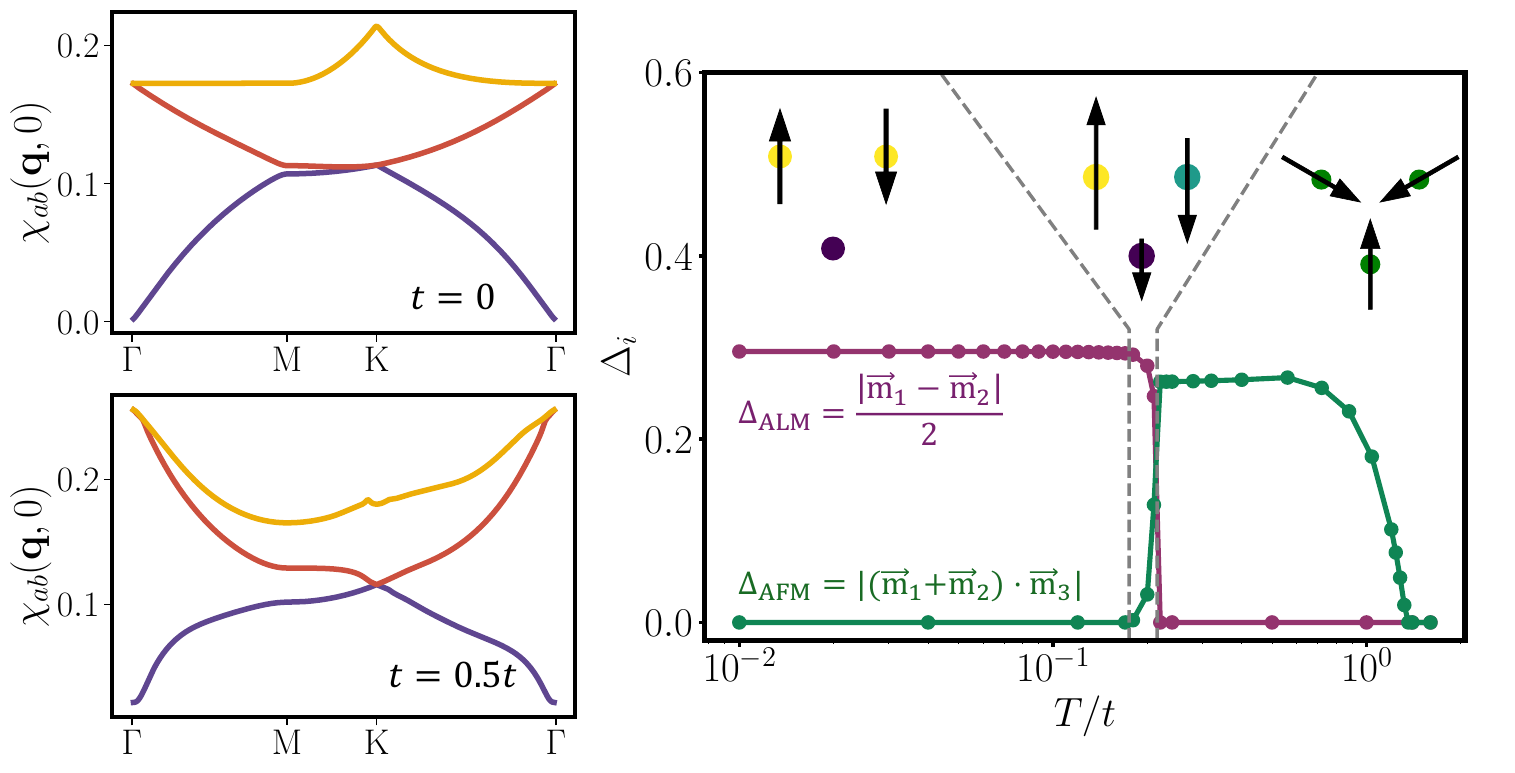}
\caption{\label{Tline} Left panels: momentum dependence of the eigenvalues of the non-interacting susceptibility matrix in the orbital basis $\chi_{ab}(\q,0)$ at $n=2/3$ and $T=0.01t$, for $t'$ indicated in each figure. Right panel: evolution of the ALM and AFM order parameters with temperature, for the HF ground state at $U=8t$, $t'=0.7t$, $n=2/3$ in a $12\times12$ lattice. The corresponding charge-spin configurations of the unit cell are sketched in each temperature interval.}
\end{figure}

\emph{Stability of the ALM state.}
We now explore the robustness of the ALM state induced by $U$ on the Kagome lattice. As we have discussed above a key ingredient for the ALM to emerge is a non-zero $t' \gtrsim 0.15 t$. 
We explore the sensitiveness of the ALM to $t'$ 
from the behavior of the non-interacting charge susceptibility, $\chi_0$, whose eigenvalues along symmetry paths of the 1BZ are shown in Figure \ref{Tline}. 
At $t'=0$, the largest eigenvalue peaks at the $K$-point
indicating an instability towards $\sqrt{3}\times\sqrt{3}$ CDW order which coincides with the periodicity of the PMDI found in HF.
However, when $t'=0.5$ the peak of the dominant eigenvalue shifts to the $\Gamma$-point, and thus we expect that $\q=(0,0)$ order is favored under $U$. The increase of the susceptibility at the $\Gamma$-point is related to the shift of the flat band occurring at high energies for $t'=0$ to the energies close to the Fermi energy, $\epsilon_F$, for non-zero $t'$ (see Supplementary Material). Hence, the large enhancement of the DOS at $\epsilon_F$ associated with the flat band favors $\q=(0,0)$ order. Among the different possible intra unit-cell spin arrangements consistent with $\q=(0,0)$ order, AFM order of the two spins is most favorable since, as discussed above, ALM order opens a gap lowering the energy of the system. This counterintuitive situation in which the doped Hubbard model 
is insulating is reminiscent of stripes found in the Hubbard model on square and ladder geometries 
which become insulating whenever the wavelength of the stripe modulation ($\lambda$) is commensurate with the hole doping ($\delta=1/\lambda$) \cite{Tocchio_stripes_2019, Metzner2023}. 

The Kagome altermagnet is not only robust under variations of $U$ and $t' \gtrsim 0.15 t$ but also upon increasing $T$. In Figure \ref{Tline} (right panel), we present the evolution of the ALM order parameter, $\D_{ALM}=|\vec{m}_1-\vec{m}_2|/2$ when all three spins in the unit cell are collinear and zero otherwise, 
with $\vec{m}_{1,2}$ the two largest mean spin vectors in the unit cell. The order parameter remains constant up to temperatures of $T\sim0.17t$, above which the magnitudes of the three mean spin vectors in the unit cell vary continuously while remaining collinear for a short temperature interval
(these intermediate states might be due to small finite-size effects). Above $T\sim0.22t$ the ground state is a $\q=(0,0)$ AFM configuration with three equal spins forming $120^{\circ}$ consistent with previous findings\cite{Kim_2020,MacDonald_AHE_2014, Tomeno1999}. The phase transition at $T\sim0.22t$ is likely first-order, since the collinear ALM cannot continuously be deformed into the coplanar AFM. Further increasing $T$, the AFM order parameter  ($\D_{AFM}\equiv|(\vec{m}_1+\vec{m}_2)\cdot\vec{m}_3|$) falls to zero continuously and the system becomes paramagnetic at $T\sim1.34t$ via a second-order phase transition.



\emph{Generalization to other lattices with larger unit cells.}
The ALMI found is not exclusive of the Kagome Hubbard model. Indeed, we have found how such configuration - or analogous realizations - is the mean field ground state of the Hubbard model on several related lattices hosting an odd number of sites in the unit cell and even integer electron filling. For instance, it emerges in the Lieb, 5-Lieb and 7-Lieb lattices \cite{ZHANG2017}(with 3, 5 and 7 sites per unit cell, respectively), where the zero-spin site is connected to the finite-spin sites by directions related to each other by a $90^{\circ}$ rotation, and
thus $|\epsilon_{AB}|\neq|\epsilon_{AC}|$ so, by the arguments given above, spin-splitting of the bands occurs. Altermagnetism also emerges in the Hubbard model in the 9-site unit cell of the Lieb-Kagome lattice \cite{Powell2025} at $n=2/9$, where spins order as three separate copies of the Kagome ALM (see supplementary). 

\begin{figure}
\includegraphics[width=\columnwidth]{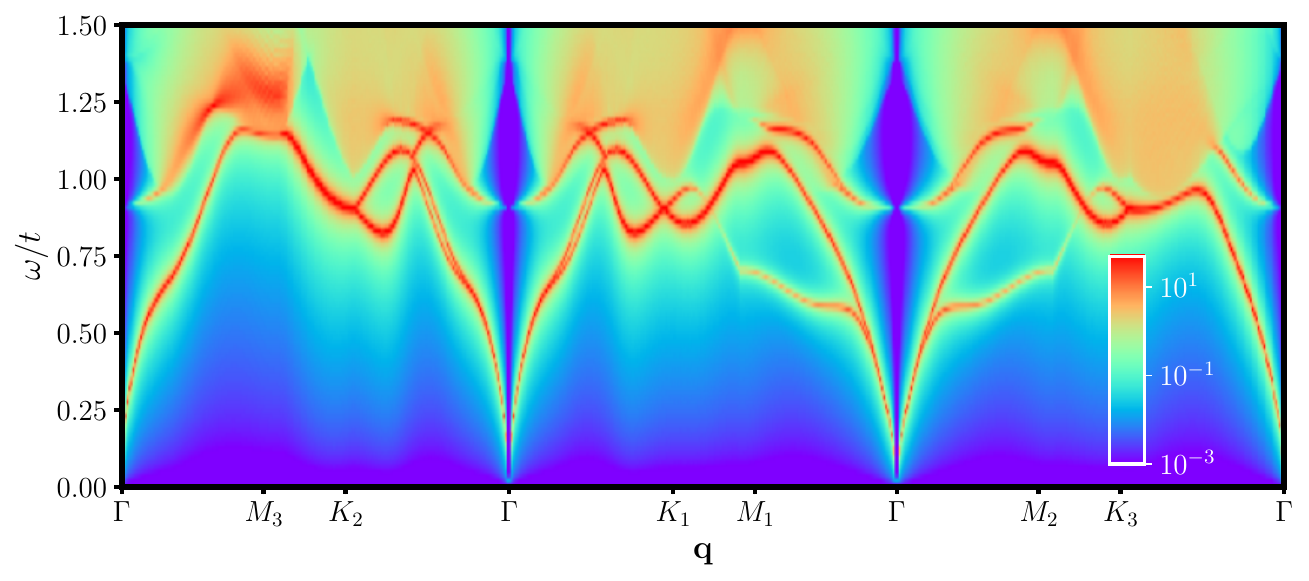}
\caption{\label{chi_KGK} Imaginary part of the physical spin susceptibility $(\chi_{+-}+\chi_{-+})(\q,\omega)/2$ of the Kagome altermagnet ($U=8t$, $t'=0.7t$, $n=2/3$, $T=0.01t$) along a momentum path connecting high-symmetry points.}
\end{figure}

Nevertheless, not all ground states with zero net magnetic moment and some spin-zero sites  are necessarily altermagnets. Some examples include the Dice and super-honeycomb lattices \cite{Sutherland1986,Zhong_2017}, as well as the triangular lattice with a $\sqrt{3}\times\sqrt{3}$ CDW and SDW. In these cases, sites $B$ and $C$ (finite spins) are related by an inversion with respect to $A$ (zero spin). This leads to $\epsilon_{AB}=\epsilon_{AC}\cc$, and bands are doubly degenerate due to the lack of $\mathcal{PT}$-symmetry breaking (see Supplementary Material).

\emph{Magnetic excitations in altermagnetic state.} 
Chiral magnons\cite{Sinova_chiral_2023} emerging in ALMs generally lead to magnon splittings which can be detected through INS \cite{Masuda2024} experiments. Our ALM occurs at moderate-to-large $U$ disappearing for $U>>t$ where it eventually becomes a QSL when $0<t'/t<1$ (see Supplementary Material). 
Thus, magnons emerging in our ALM are well captured at the mean-field level. We extract the imaginary part of the spin susceptibility, $\Im\chi_{ij}(\q,\omega)$, relevant to INS using a multiorbital RPA approach \cite{Maier2023} where $\chi_{ij}(\q,\omega)$ reads: 

\begin{equation}
    \chi_{+-}(\q,\omega)=\sum_{l,l'}(\chi_{RPA}(\q,\omega))^{l\up,l\down}_{l'\up,l'\down} .
\end{equation}
Here, $l,l'$ run over the orbitals, the quantities in the r.h.s. are the elements of the spin susceptibility tensor in the multiorbital RPA, and $\chi_{-+}(\q,\omega)$ is obtained by exchanging $\up\longleftrightarrow \down$. The RPA is known to be adequate at small-to-moderate $U$.

In Fig. \ref{chi_KGK}, we show the averaged spin-flip spectra, $\Im(\chi_{+-}(\q,\omega) + \chi_{-+}(\q,\omega) )/2$, along a path connecting high-symmetry points in the 1BZ shown in Figure \ref{fig:Kagome_altermagnet}b). Due to the spin-splitting of the ALM bands (see Fig. \ref{fig:Kagome_altermagnet}c)), the INS signal at transferred momenta, $\q$, corresponding to spin-flip transitions between occupied and empty spin-splitted bands will readily depend on the chirality of the process. Hence, the magnon spectra displays chirality-splitted magnon bands only at the ${\bf q}$  
at which the ALM bandstructure is spin-splitted ({\it e.g.}, along the  $M_1\rightarrow\Gamma\rightarrow M_2$ path). 
The origin of the splitting in our magnon bands differs from the one found in the modified Lieb lattice Hubbard model \cite{franz_prl_2025} at large $U$. In such
``anti-CuO$_2$" model, ALM is due to anisotropies in the n.n.n. spin exchange couplings associated with the presence of spinless O-sites which are absent in our spatially uniform  Kagome and Lieb Hubbard models. 

\emph{Conclusions.}
In this work we report an ALM phase arising in the single-band
Kagome Hubbard model induced solely by $U$. The collinear ALMI breaking $\mathcal{PT}$-symmetry is also relevant to the Hubbard model on Lieb and Kagome lattices with large unit cells whose common feature is that they consist of an odd number of sites occupied by an even number of electrons. Since the ALM is robust in a broad range of parameters and temperatures,
chiral magnons could be detected through INS on interacting Kagome systems. 

\acknowledgments
We acknowledge financial support from (Grant No. PID2022-139995NB-I00) MICIN/FEDER, Uni\'on Europea and
from the Mar\'ia de Maeztu Programme for Units of Excellence in R\&D (Grant No. CEX2023-001316-M). A. B. acknowledges financial support from the FPU Grant No. FPU23/02149.

\bibliography{bibliography}

\end{document}


\title{Supplementary Material for Altermagnetism in an interacting model for Kagome materials}

\author{Alejandro Blanco Peces}
\email[]{alejandro.blancop@uam.es}
\author{Jaime Merino}
\email[]{jaime.merino@uam.es}
\affiliation{Departamento de Física Teórica de la Materia Condensada, Condensed Matter Physics Center, (IFIMAC) and Instituto Nicolás Cabrera, Universidad Autónoma de Madrid, 28049 Madrid, Spain}
\date{\today}
\begin{abstract}
\end{abstract}

\maketitle

\section{Hartree-Fock approximation at finite temperatures}
 
 In this work, we use a real space Hartree-Fock (HF) approach to determine the ground state of the Hubbard model on the Kagome lattice of the main text. Following Ref. \cite{Metzner2023}, we use a finite size lattice with $N_{a,b}$ cells along the lattice directions with periodic boundary conditions (PBC). We calculate the properties of the model for different filling fractions, temperatures, and $t'/t$ ratios. The model reads:
\begin{equation}
    \H=\sum_{ij}t_{ij}c\d_{i\sigma}c_{j\sigma}+U\sum_j n_{j\up}n_{j\down}
\end{equation}
where $c_{i\sigma}$ ($c\d_{i\sigma}$) annihilates (creates) an electron at site $i$ with spin $\sigma=\up,\down$,
$t_{ij}$ is the hopping element between sites $ij$, $U$ is the onsite Hubbard repulsion and $n_{i\sigma}=c\d_{i\sigma}c_{i\sigma}$ is the electron filling of site $i$. In the following, we set $t_{ij}=-t$ if $ij$ are first neighbors and $t_{ij}=-t'$ if $ij$ are second neighbors. We perform a mean-field decoupling of the Hubbard term: 
\begin{multline}
    \H_U^{MF}=U\sum_i \Delta_{i\ol{\sigma}}n_{i\sigma}-U\sum_i(\Delta_{i+}c\d_{i\down}c_{i\up}+h.c.) \\
    +U\sum_i(\Delta_{j\up}\Delta_{j\down}-\Delta_{j-}\Delta_{j+}) \, 
\end{multline}
with $\ol{\sigma}\equiv -\sigma$, where we have defined 
\begin{align}                                   &\Delta_{j\sigma}=\expval{n_{j\sigma}} \\
    & \Delta_{j+}=\Delta_{j-}\cc=\expval{c\d_{i\up}c_{i\down}}
\end{align}
which are determined self-consistently to obtain the ground state of the system. The last term in the mean field Hamiltonian is a constant contribution, which we will drop from now on but we will take into account when comparing the free energies of various competing ground states. The minimization of the HF energy for given parameters: $U$, $t$, $n$, $T$ is achieved through an iterative procedure. In a first iteration, we construct the Hamiltonian matrix assuming a random set or a physically reasonable {\it Ansatz}  $\{\Delta_{j\up},\Delta_{j\down},\Re{\Delta_{j+}},\Im{\Delta_{j+}}\}$, diagonalizing the HF hamiltonian to obtain the eigenenergies, $\epsilon_l$, and eigenvectors, $v^l_{i\sigma}$. At each step, the chemical potential $\mu$ is determined by fixing the total number of electrons in the lattice $n_e=n\times N_{s}$: 
\begin{equation}
    n=\frac{1}{N_s}\sum_{i,\sigma}\langle n_{i\sigma}\rangle=\frac{1}{N_s}\sum_{l=1}^{2N_s}f(\epsilon_l-\mu),
\end{equation}
where $N_{s}=3\times N_a\times N_b$ is the number of sites in the finite lattice and $f(\epsilon)$ is the Fermi-Dirac distribution. Note that this procedure requires the filling fraction to be a rational number. At zero temperature, $\mu$ is easily found by filling the lowest $n_e$ states. Then, the new set of variational parameters is calculated by taking mean values of the Hamiltonian matrix:
\begin{equation}
    \langle c^\dagger_{i\sigma}c_{i\sigma'}\rangle=\sum_{l=1}^{2N_s} (v^l_{i\sigma})\cc v^l_{i\sigma'} f(\epsilon_l-\mu) 
\end{equation}
At $T=0$, the Fermi function reduces to the Heaviside theta function so the sum over $l$  only runs over the lowest $n_e$ eigenvalues of the Hamiltonian. In a second iteration the Hamiltonian is updated, typically mixing the new and the old variational parameter sets to accelerate the convergence. Such iterative process is repeated until the differences between variational parameters in two consecutive steps do not exceed a certain tolerance threshold $\delta \sim 10^{-15}$. The general iterative procedure described is used to determine the ground state of other two-dimensional lattices with larger unit cells discussed in the main text.

One may encounter situations in which the ground state found within such real space HF approach is spurious due to finite-size effects. The stability of such ground states may be checked by increasing the size of the lattice but this greatly increases the computational cost since the time to diagonalize the matrix scales as $N_s^3$, and larger lattices need a larger number of iterations to converge. Nevertheless, the altermagnetic configurations found do not break the translational symmetry of the lattice, so that it is straightforward to check their stability in the thermodynamic limit. So in order to do this we express the H-F equations in momentum by taking the Fourier transform of the Hubbard Hamiltonian:
\begin{equation}
    \H=\frac{1}{N_k}\sum_{\k}\sum_{ij}t_{ij}c\d_{i\sigma\k}c_{j\sigma\k}e^{i\k\cdot(\r_i-\r_j)}+U\sum_j n_{j\up}n_{j\down}
\end{equation}
where now $i$ runs over the sites of a single unit cell of the original lattice which are located at positions $\r_i$. $N_k$ is the number of momentum points $\k$ assumed in the  discretized 1BZ. Starting from a random set of variational parameters, we determine the ground state of the lattice self-consistently summing over the discretized momenta, the sites in the unit cell (three for the Kagome lattice) and the two spins per site. 

\section{Analysis of spontaneously broken symmetry ground states}
Based on the real space HF procedure, we obtain the ground state of the system on a finite lattice which is fully determined by the set of parameters $\{\Delta_{j\up},\Delta_{j\down},\Re{\Delta_{j+}},\Im{\Delta_{j+}}\}$. These parameters are related to the mean electron filling and spin at the lattice sites by:
\begin{align}
     &\langle n_i \rangle = \Delta_{i\up}+\Delta_{i\down} \\
    & \langle S^z_i \rangle = \frac{1}{2}(\Delta_{i\up}-\Delta_{i\down}) \\
    & \langle S^x_i \rangle = \frac{1}{2}(\Delta_{i+}+\Delta_{i-}) = \Re{\Delta_{i+}} \\
    & \langle S^y_i \rangle = \frac{1}{2i}(\Delta_{i+}-\Delta_{i-}) = \Im{\Delta_{i+}}.
\end{align}

By including the spin-flip terms containing $\D_{i\pm}$ in the mean-field Hamiltonian, we can find ground states where the spin configurations are collinear, coplanar and non-coplanar. However, the ALM found turns out to be collinear. If we allow the spins to point in any direction, the axis along which the non-zero spins are aligned is randomly selected by the iterative process, and it is equivalent to dropping the $\D_{i\pm}$ terms of the Hamiltonian and solving for the ground state in which the spins are directed along the $z$-axis. 

After solving for the real space charge and spin vector distributions, we perform a Fourier transform to characterize their 
possible order:
\begin{align}
    & \expval{n_{\q} }= \sum_{j} \expval{n_j } e^{i\q\cdot\r_j} \\
    & \expval{\boldsymbol{S}_{\q} }= \sum_{j} \expval{\boldsymbol{S}_j } e^{i\q\cdot\r_j},
\end{align}
where the sum runs over all sites $j$ of the lattice and $\r_j$ are their real space positions. 

Since the ALM patterns found consist of a single unit cell repeated over the lattice they preserve the translation symmetry of the Kagome lattice. Therefore, the variational parameters corresponding to our ALM satisfy the following conditions:
\begin{itemize}
    \item The Fourier coefficients $\expval{n_{\q}}$ and $\expval{\boldsymbol{S}_{\q}}$ are finite only for $\q=(0,0)$ and zero otherwise, indicating the absence of charge and spin modulations with an enlarged unit cell of the Kagome lattice.
    \item The sum of the three spins in a unit cell (all cells are equivalent) of the Kagome lattice vanishes, since a zero net magnetic moment is a requirement for altermagnetism.
    \item One of the three spin vectors in the unit cell of the Kagome lattice vanishes, as other AFM configurations are consistent with the above conditions.
\end{itemize}

When only the first two conditions are satisfied, the ground state is labeled as an antiferromagnet (AFM) if the spins have finite size, or as a $120^{\circ}$AFM if all three spins have the same magnitude but they form $120^{\circ}$ with respect to each other. If all three spins are zero, the state is a paramagnet (PM). This is the state found at sufficiently high $T$ or for sufficiently weak $U$, as expected. The last possibility of a ${\bf q}=(0,0)$ ground state has a net magnetic moment in the unit cell so that it is labeled a ferromagnet (FM). 

Aside from the particular set of $\q=(0,0)$ orders discussed above, the Hartree-Fock ground state of the Kagome Hubbard model can be much more complex involving Fourier coefficients with $\q \neq (0,0)$. We have identified periodic charge and spin modulations on the Kagome lattice (see Fig. 1 of main text), dubbed Charge (CDW) and Spin Density Wave (SDW) from the $\bf{q}$ dependence of the Fourier coefficients. 
Hence, all the periodic density wave (DW) patterns encountered in our calculations can be classified as follows: 

\begin{itemize}
    \item A $2\times2$ DW order which has non-vanishing Fourier coefficients (aside from $\q=(0,0)$) at $\q_i=\boldsymbol{g}_i/2$ and linear combinations of them within the 1st Brillouin Zone.
    \item A $\sqrt{3}\times\sqrt{3}$ DW order with $\q_{1}=\boldsymbol{g}_1/3-\boldsymbol{g}_2/3$ and $\q_{2}=\boldsymbol{g}_1/3+2\boldsymbol{g}_2/3$;
    \item A $2\sqrt{3}\times2\sqrt{3}$ with $\q_{1}=\boldsymbol{g}_1/6-\boldsymbol{g}_2/6$ and $\q_{2}=\boldsymbol{g}_1/6+\boldsymbol{g}_2/3$. 
\end{itemize}

CDW and SDWs can coexist within the ground state and even have different periodicities. We have also found charge and spin striped patterns. At large $U$, where the Hartree-Fock approximation breaks down we find disordered phases which may be associated with quantum spin liquid (QSL) behavior. The occurrence of a QSL in our model is supported by the ED calculations described below in App. \ref{app:F}.

\section{Altermagnetic ground states}

\begin{figure*}[t!]
\centering
{\includegraphics[width=\textwidth]{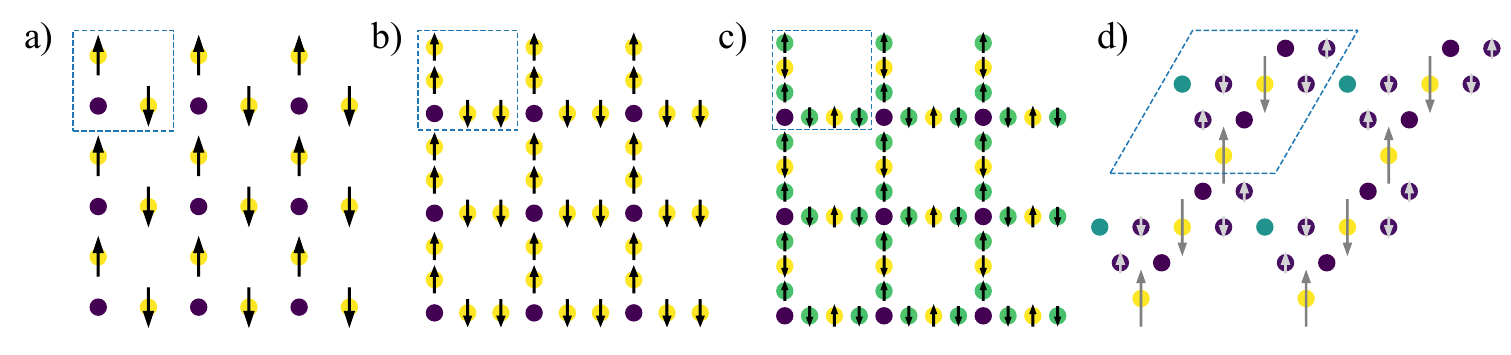}}
\caption{Real space altermagnetic patterns obtained as the mean-field ground states of the Hubbard model on other lattices different from the Kagome: a) Lieb lattice, b) 5-Lieb lattice, c) 7-Lieb lattice, d) Lieb-Kagome lattice. In each case, yellow (purple) circles correspond to sites with the highest (lowest) charge density and blue-green colors correspond to intermediate densities. Arrows indicate the directions of the average spin vectors, $\expval{\Vec{S}_i}$, and dashed blue lines enclose the unit cell of each lattice. Sites without an arrow correspond to $\expval{\Vec{S}_i}=0$.
}
\label{other_altermagnets}
\end{figure*}

\begin{figure}
\centering
\includegraphics[width=\columnwidth]{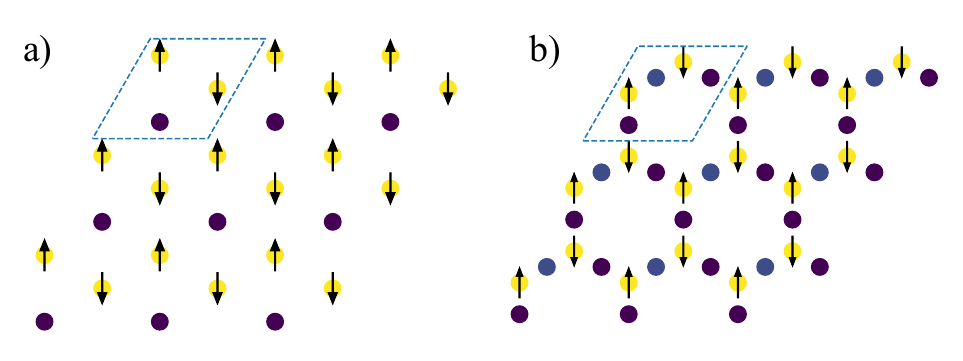}
\caption{\label{non_altermagnets}Real space patterns obtained as ground states of the Hubbard model on the a) Dice/triangular lattice, b) super-honeycomb lattice that are not altermagnets. These states consist of unit cells with zero and finite spin and zero net momentum as in Fig. \ref{other_altermagnets} but they are not altermagnets
since they preserve $PT$-symmetry.
In each case, yellow (purple) circles correspond to sites with the highest (lowest) charge density and blue-green colors correspond to intermediate densities. Arrows indicate the directions of the average spin vectors, $\expval{\vec{S}_i}$, and dashed blue lines enclose the unit cell of each lattice. Sites without an arrow 
correspond to $\expval{\vec{S}_i}=0$.
}
\end{figure}

The ALM found (see Fig. 2 a) of the main text) is not restricted to the Kagome lattice. Analogous ALM configurations arise in other lattices with an odd number of sites per unit cell and with an even number of electrons. Aside from the Kagome lattice, we have studied the Lieb, 5-Lieb, 7-Lieb, Lieb-Kagome, triangular, Dice and super-honeycomb lattices, which have either 3, 5, 7 or 9 sites per unit cell. 
The altermagnetic ground state of the Lieb lattice (Fig. \ref{other_altermagnets}a)) arises at $n=2/3$ ($n=4/3$) when a finite negative (positive) n.n.n. hopping is included, in agreement with \cite{franz_prl_2025,thomale_prl_2025}, with the difference that we do not include a site-dependent onsite energy and treat all sites as magnetic. Here, the zero-spin site is connected to the finite-spin sites by directions related to each other by a 90$^{\circ}$ rotation, and thus $|\e_{AB}| \neq |\e_{AC}|$ so, by the arguments given in the main text, spin-splitting of the bands occurs. Further decorations of the Hubbard model include the 5-Lieb and 7-Lieb lattices, which place 2 and 3 extra sites, respectively, equally spaced in each bond of the square lattice. We have found these lattices to have ALM ground states in our mean field treatment, shown in Fig. \ref{other_altermagnets}b-c). While the 5-Lieb lattice needs second- and third-neighbor hopping at $n=2/5$ or $n=8/5$ to support the ALM state, in the 7-Lieb lattice at $n=6/7$ a finite n.n.n. suffices. Altermagnetism also emerges in the Hubbard model on the Lieb-Kagome lattice, which can be viewed as a decorated Kagome lattice featuring one extra site at the middle of each bond and nine sites per unit cell. At $n=2/9$ an altermagnet appears as the ground state, spins order as three separate copies of the Kagome altermagnet (see Fig. \ref{other_altermagnets}d)). Although the growing number of sites per unit cell increases the complexity of the ground state, they have in common that not only the sum of magnetic moments adds up to zero, but there is also some symmetry breaking with respect to the original lattice: sites that were originally equivalent have now opposite spins. It can be easily checked that all the depicted configurations break $\mathcal{PT}$-symmetry: by flipping all spins followed by an inversion of coordinates with respect to an empty site, the final configuration is
different from the initial one. The absence of $\mathcal{PT}$-symmetry is a characteristic feature of ALMs associated with the breaking of Kramers degeneracy. The original $C_n$ symmetry of the lattices ($n=6$ for the Lieb-Kagome lattice and $n=4$ for the rest) is also broken down to $C_2$ symmetry, and the spin character of their band structures shows this, as in the Kagome lattice case. 

Note that a ground state showing $\q=(0,0)$ order with some empty sites and spins adding up to zero is not necessarily an altermagnet. It needs to explicitly break $PT$-symmetry, or equivalently, the Hamiltonian matrix elements in momentum space should satisfy: $|\epsilon_{AB}|\neq|\epsilon_{AC}|$, with $A$ being the spin zero site and $B$, $C$ having opposite finite spins. For instance, the mean-field ground state of the triangular lattice (one-site basis) with n.n. hopping and $n=2/3$ is a $\sqrt{3}\times\sqrt{3}$ CDW and SDW with an enlarged three-site unit cell shown in Figure \ref{non_altermagnets}a). The Dice lattice (a three-site basis lattice that comes from removing one third of the bonds of the triangular lattice) has the same ground state at $n=2/3$. Although they resemble the Kagome altermagnet, sites $B$ and $C$ are related by an inversion with respect to A. This leads to $\epsilon_{AB}=\epsilon_{AC}\cc$, and the bands end up being doubly degenerate. This behavior also appears in the super-honeycomb lattice (a graphene lattice decorated with an extra site in the midpoint of each bond, with five sites per unit cell). With two electrons per unit cell, the three bond sites have zero spin, while the pair of original graphene
sites have opposite spins (see Fig. \ref{non_altermagnets}b)), but there is no spin splitting since the system is $\mathcal{PT}$-symmetric.

\section{Analysis of tight-binding band structure}

\begin{figure*}[t!]
\centering
{\includegraphics[width=\textwidth]{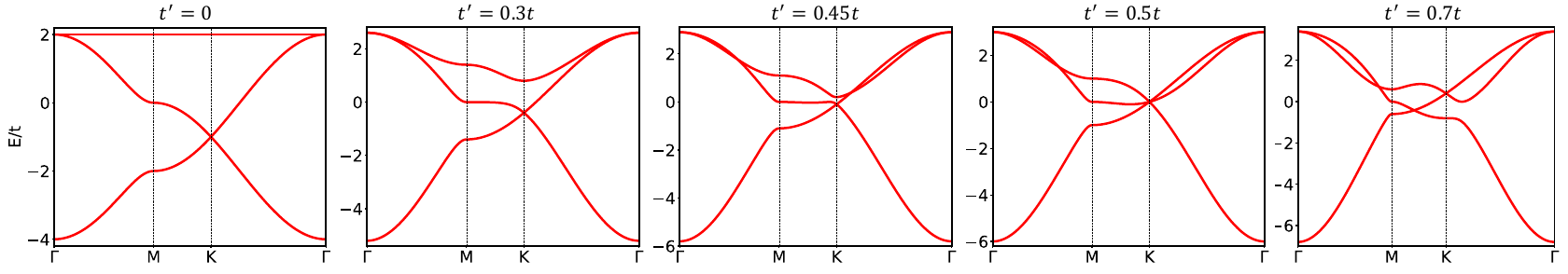}}
\caption{Dependence of the tight-binding band structure of the Kagome lattice with $t'$.}
\label{band_evol}
\end{figure*}
We now discuss the dependence of the tight-binding band structure on $t'/t$. At $U=0$ our model reduces to a tight-binding model which, in momentum space, reads: 
\begin{equation*}
    \mathcal{H}_0=\sum_{\boldsymbol{k}}\sum_{i,j,\s}c^{\dagger}_{i\sigma\k}(H_{\s}(\boldsymbol{k}))_{ij}c_{j\sigma\k},
\end{equation*}
where the Hamiltonian matrix in the site basis reads:
\begin{equation}\label{H0}
    H_{\s}(\k)=\begin{pmatrix}
        0 & \e_{AB} & \e_{AC} \\
        \e_{AB} & 0 & \e_{BC} \\
        \e_{AC} & \e_{BC} & 0 
    \end{pmatrix}.
\end{equation}

Here, we have used that $\e_{ij}=\e_{ji}$ since all matrix elements are real:
\begin{align*}
    \e_{AB}=-2t\cos(\k\cdot\boldsymbol{a_2}/2)-2t'\cos(\k\cdot(\boldsymbol{a_3}-\boldsymbol{a_1})/2) \\
    \e_{AC}=-2t\cos(\k\cdot\boldsymbol{a_3}/2)-2t'\cos(\k\cdot(\boldsymbol{a_2}+\boldsymbol{a_1})/2) \\
    \e_{BC}=-2t\cos(\k\cdot\boldsymbol{a_1}/2)-2t'\cos(\k\cdot(\boldsymbol{a_3}+\boldsymbol{a_2})/2)
\end{align*}
where $\boldsymbol{a_1}=(1,0)$, $\boldsymbol{a_2}=(1/2, \sqrt{3}/2)$, $\boldsymbol{a_3}=\boldsymbol{a_2}-\boldsymbol{a_1}$ are the real-space lattice vectors of the Kagome lattice. $t$ and $t'$ are the n.n. and n.n.n hopping amplitudes, respectively. For $t'=0$, this matrix can be analytically diagonalized yielding:
\begin{align*}
    &E_1=2t \\
    &E_{2,3}=-t\left(1\pm\sqrt{4(c^2_1+c^2_2+c^3_3)-3}\right)  
\end{align*}
with $c_i\equiv \cos(\k\cdot\boldsymbol{a}_i/2)$. That is, the band structure consists on two dispersive bands and a flat band. The dispersive bands touch the flat band at a quadratic band crossing (QBC) occurring at the $\Gamma$ point, they cross each other at Dirac cones located at the $K$ points, and the density of states diverges at the $M$ points due to the presence of Van Hove singularities. 

When $t' \neq 0$, the flat band becomes dispersive and may cross the other bands, producing new Dirac cones. For increasing $t'$, one of these Dirac cones approaches the $K$ point, and at $t'=0.5t$ all bands cross at the $K$-point. 
At even larger $t'$, the lowest Dirac cone is displaced along the $M-K$ path, while the Dirac cone at $K$ remains, but it is produced now by a crossing of the two upper bands. This band evolution is illustrated in Figure \ref{band_evol}. Thus, even if the Dirac point at $K$ is protected (if the little group $C_{3v}$ at this point has a two-dimensional irreducible representation that forces a band degeneracy), it is not the breaking of such symmetry which makes the ALM  insulating, since for $t'>0.5t$ the gap opening at the Fermi level is independent of the band crossing at $K$.

\section{Origin of altermagnetic gap opening}

The eigenvalue equation of the original tight-binding Hamiltonian \eqref{H0} is
\begin{equation}
    E^3-E(\e_{AB}^2+\e_{BC}^2+\e_{AC}^2)-2\e_{AB}\e_{BC}\e_{AC}=0
\end{equation}
which is a depressed cubic equation of the form $E^3+pE+q=0$. For the bands to be degenerate (QBC, triple crossing or Dirac cone) at a given momentum, $\k$, the discriminant $\Delta=-4p^3-27q^2$ must vanish:
\begin{equation}
    (\e_{AB}^2+\e_{AC}^2+\e_{BC}^2)^3=27\e_{AB}^2\e_{BC}^2\e_{AC}^2.
\end{equation}
However, the arithmetic mean-geometric mean (AM-GM) inequality leads to $(\e_{AB}^2+\e_{AC}^2+\e_{BC}^2)^3\geq27\e_{AB}^2\e_{BC}^2\e_{AC}^2$, where the equality holds for $|\e_{AB}|=|\e_{BC}|=|\e_{AC}|$. Therefore, there will be a band crossing in the non-interacting system at those momenta $\k$ at which $|\e_{AB}|=|\e_{BC}|=|\e_{AC}|$. Defining $|\e_{AB}|=|\e_{BC}|=|\e_{AC}|\equiv t$ and sgn$(\e_{AB}\e_{BC}\e_{AC})\equiv s$, the eigenvalue equation
\begin{equation*}
    E^3-3Et^2-2st^3=0
\end{equation*}
has a simple root $E_1=2st$ and a double root $E_2=E_3=-st$. At $t=0$, the simple and double roots merge and become a triple root.

Introducing the altermagnet minimal model with a site-dependent Zeeman field, the Hamiltonian matrix for the sector with spin $\s=\{+1,-1\}$ changes to:
\begin{equation}\label{H0ALM}
    H_{\s}(\k)=\begin{pmatrix}
        0 & \e_{AB} & \e_{AC} \\
        \e_{AB} & J\s & \e_{BC} \\
        \e_{AC} & \e_{BC} & -J\s 
    \end{pmatrix}
\end{equation}
whose eigenvalue equation reads:
\begin{multline}
    E^3-EJ^2-E(\e_{AB}^2+\e_{BC}^2+\e_{AC}^2)\\
    -2\e_{AB}\e_{BC}\e_{AC}-J\s(\e_{AB}^2-\e_{AC}^2)=0.
\end{multline}

But at the momenta where there was initially a degeneracy ($|\e_{AB}|=|\e_{BC}|=|\e_{AC}|$) the above equation simplifies to:
\begin{equation*}
    E^3-E(3t^2+J^2)-2st^3=0,
\end{equation*}
and it is straightforward that the discriminant does not vanish anymore:
\begin{equation*}
    \Delta=4J^2(27t^4+9t^2J^2+J^4) >0 \text{, if } J\neq0 
\end{equation*}
There are then three distinct solutions and all previous degeneracies (not only Dirac cones at the Fermi level) are broken. Note also that the eigenvalue equation is invariant under the transformation $\s\rightarrow-\s$, so the spin-up and spin-down eigenenergies are identical. Thus, although the Dirac cone is gapped, the spin-up and spin-down bands will cross at these momenta, while at the momenta at which $|\e_{AB}|\neq|\e_{AC}|$ we have already shown that they become non-degenerate. 

Let us now determine the size of the gap for a small Zeeman field $J$, and define $\varepsilon=J^2$. Consider the function 
\begin{equation}
    f(E,\varepsilon)=E^3-2st^3-3Et^2-E\varepsilon,
\end{equation}

Now, the energy eigenvalues at momenta with a band degeneracy will be given by the solutions of $f(E,\ve)=0$. For $J=0$, we know that the function factorizes as:
\begin{equation}
    f(E,0)=(E-2st)(E+st)^2.
\end{equation}

Taking $E_0=-st$ (double root) and expanding $f(E,\ve)$ around $f(E_0,0)$, with $E=E_0+\delta$:
\begin{equation}
    f(E,\ve)\approx \frac{\partial f}{\partial \ve}\bigg\vert_{E_0,0}\ve+\frac{1}{2}\frac{\partial^2 f}{\partial E^2}\bigg\vert_{E_0,0}\delta^2 
\end{equation}
as both $f$ and its partial derivative with respect to $E$ vanish at $(E_0,0)$. Since $f_{\ve}(E_0,0)=st$ and $f_{EE}(E_0,0)=-6st$, imposing that the Taylor expansion above vanishes leads to $\delta^2=\ve/3$. Then, the degenerate solution $E_0$ splits into
\begin{equation}
    E_{\pm}=E_0\pm\frac{|J|}{\sqrt{3}}
\end{equation}
and the size of the gap is 
\begin{equation}
    2\delta=\frac{2|J|}{\sqrt{3}}
\end{equation}

The case of a triple root is simpler: it occurs when $t=0$, and the eigenvalue equation reads $E^3=0$. The triply degenerate eigenvalue is $E_0=0$. Upon introducing the effective Zeeman field $J$ to model the altermagnet, the new eigenvalue equation becomes $E(E^2-J^2)=0$, with solutions $E_0=0$ and $E_{\pm}=\pm|J|$. Again, a gap is opened between all previously crossing bands with a gap that depends linearly 
on $|J|$.

Thus, all band degeneracies of the non-interacting system break when introducing our  minimal model for the altermagnetic state on the Kagome lattice with a gap increasing linearly with $|J|$ at small $|J|$. Moreover, this discussion is general, without using explicitly the hopping amplitudes $\e_{ij}(\k)$ on the Kagome lattice. Therefore, it accounts as well for the gap opening in other three-site lattices with analogous spin patterns (zero net magnetic moment and one spin zero site in the unit cell) regardless of whether altermagnetism is present or not. 
Hence, the ALM found in the Lieb lattice and the AFM states occurring in the Dice/triangular lattices are insulators even though spin-splitting does not occur in the latter cases.
A key feature underlying the insulating character of these states is that their unit cell contains an even number of electrons.


\begin{figure*}[t!]
\centering
\includegraphics[width=\textwidth]{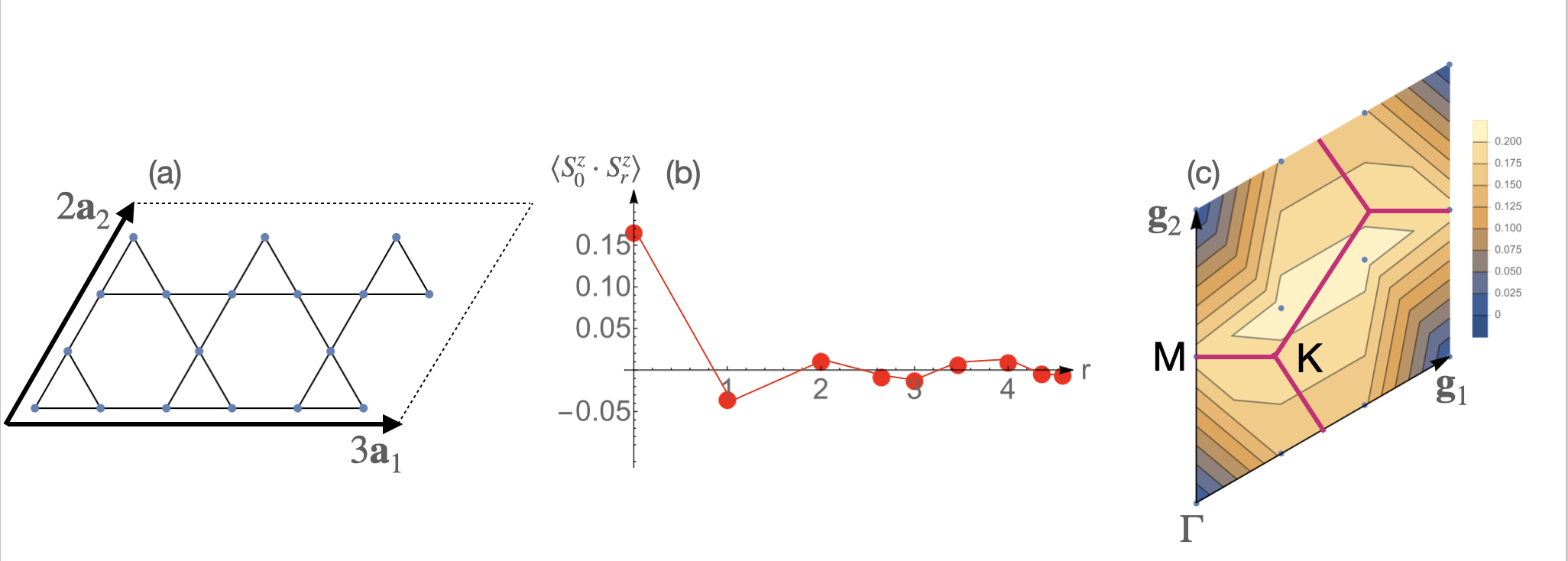}
\caption{Spin correlations in the Hubbard model on the Kagome lattice for $U=20$ and $t'=\frac{1}{\sqrt 2}$ from ED calculations. (a) The $N_s=18$ site cluster with PBC used in the ED calculation, (b) the dependence of the spin correlations with relative distance, $\langle  S^z_0 \cdot S^z_i\rangle$, (c) the static spin structure factor $S({\bf q})$  at momenta in the region enclosed by the two primitive vectors of the reciprocal lattice ${\bf g}_1$ and ${\bf g}_2$, including the $\Gamma$, $K$ and $M$ points. The blue circles are the ${\bf q} $ vectors allowed by the PBC on the cluster. The solid (pink) lines denote the 1BZ.}
\label{fig:SqU20}
\end{figure*}

\section{Exact diagonalization on small clusters}
\label{app:F}
Based on the H-F calculations of the main text, while the difference in charge density between filled and empty sites first grows as $U$ increases, at strong enough $U$, a uniform charge distribution is preferred. Thus, at $U>>t$, the ALM should give way to a strongly correlated uniform metallic state and the system should be regarded as a highly doped Mott insulator ($\delta=1/3$). 
In order to explore electron correlation effects in this limit, we have performed 
exact diagonalization (ED) calculations of the Hubbard model on a Kagome lattice on the small $N_s=3 \times 3\times 2$ clusters with PBC shown in Fig \ref{fig:SqU20}(a) for $n=2/3$. We have obtained the real space spin correlations, $\langle S_0^z \cdot S_r^z \rangle $ where $r$ is the relative distance between the reference site $0$ and any other site in the cluster. As shown in Fig. \ref{fig:SqU20} (b) a rapid suppression of $\langle S_0^z \cdot S_r^z \rangle $ with $r$ is found suggesting the presence of a spin disordered QSL state. 

In order to explore the spin pattern of possible ordered phases, we have also calculated the static spin structure factor which reads:
\begin{equation}
S^{zz}({\bf q}) = \frac{1}{N_s} \sum_{i,j} e^{i {\bf q} ({\bf r}_i -{\bf r}_j) } \langle S^z_i S^z_j \rangle,
\end{equation}
where $i,j$ are lattice sites in the cluster. 

In Fig. \ref{fig:SqU20} (c) we show $S^{zz}({\bf q})$ for $U=20$. The spin structure factor is quite structureless with no characteristic peaks which could indicate a preferred magnetic order within the  resolution allowed by our cluster. The negligible contribution from the ${\bf q} \sim 0$ to $S({\bf q})$ suggests that the altermagnetic order is not present at large $U$. On the other hand, the absence of the $K$-point in our finite cluster calculation does not allow to conclude about the possibility of $\sqrt{3}\times \sqrt{3} $ order. However, our results resemble qualitatively previous ED and DMRG calculations \cite{Sheng2008,Lauchli2009,McCullloch2015} of the Heisenberg model on the Kagome lattice where a broad but small peak at the border of the extended Brillouin zone occurs with negligible 
contribution from the $\Gamma$ point. These have been interpreted as 
signatures of a QSL. A similar structure of $S({\bf q})$ 
survives down to low $U=5$ values with an overall suppression of spectral weight from a maximum in 
$S({\bf q})$ of around 0.2 to 0.14 (see Fig. \ref{fig:SqU20}).
Based on this comparison, our ED results suggest that the most likely 
ground state of the Hubbard model on the Kagome lattice for $n=2/3$ (1/3-filling) is a magnetically disordered state, most likely, a QSL as predicted for $n=1$. Further calculations on larger systems are needed to corroborate these findings.

\section{Bare susceptibility}

In order to test the consistency of the ALM state, we have also calculated the non-interacting charge susceptibility which reads:
\begin{multline}\label{eq:chi0}
    (\chi^0)^{st}_{pq}(\q,\omega)=-\frac{1}{N}\sum_{\k,\mu,\nu}\frac{f(\e_{\nu}(\k+\q))-f(\e_{\mu}(\k))}{\omega+\e_{\nu}(\k+\q)-\e_{\mu}(\k)+i\eta}\times\\
    \times a^s_{\mu}(\k)(a^p_{\mu}(\k))\cc a^q_{\nu}(\k+\q)(a^t_{\nu} (\k+\q))\cc,
\end{multline}
where $N$ is the number of momentum points, $\k$, in the discretized  1BZ. $\mu,\nu$ run over the band indices, $\e_\mu(\k)$ are the energy eigenvalues, $f(\e)$ is the Fermi-Dirac distribution, $a^s_\mu(\k)$ are the components of the eigenvector of the Hamiltonian with eigenvalue $\e_\mu(\k)$ and $\eta$ is an infinitesimal quantity. The four indices $s,t,p,q$ of the tensor run over the orbitals. From the above tensor we construct the bare susceptibility matrix in the orbital basis:
\begin{equation}
    \chi_{sp}(\q,\omega)=\Re(\chi^0)^{ss}_{pp}(\q,\omega).
\end{equation}
We plot the eigenvalues along a path in momentum space
for $\omega=0$ (static limit) in Figure 3 of the main text.

\section{Physical spin susceptibility}

In the ALM found on the Kagome lattice, the actual INS intensity is related to the the physical spin susceptibility which is defined as: 
\begin{equation}
    \chi_{ij}(\q,\tau)=-\frac{1}{N}\sum_{ll'}\expval{\mathcal{T}_\tau S_i^l(\q,\tau)S_j^{l'}(-\q,0)},
\end{equation}
where $l,l'$ run over the orbitals, and $ij=\{+-,-+\}$. Using the spin-flip operators $S_+^l(\q,\tau)=\sum_{\k}c\d_{l\up,\k+\q}(\tau)c_{l\down,\k}(\tau)$ and $S_-^l(\q,\tau)=\sum_{\k}c\d_{l\down,\k+\q}(\tau)c_{l\up,\k}(\tau)$, Wick's theorem and replacing time-ordered expectation values by one-particle Green's functions one finds: $\expval{\mathcal{T}_\tau c_{l,\k}(\tau)c\d_{l',\k}(0)}=G^{l,l'}(\k,\tau)$:
\begin{equation}
    \chi_{+-}(\q,\tau)=-\frac{1}{N}\sum_{ll',\k}G^{l'\up,l\up}(\k+\q,-\tau)G^{l\down,l'\down}(\k,\tau),
\end{equation}
and $\chi_{-+}(\q,\tau)$ is obtained by exchanging $\up\longleftrightarrow\down$ in the r.h.s of the above equation. Performing the Fourier transform from imaginary time, $\tau$, to bosonic Matsubara frequencies, $\omega_m$, and after analytical continuation to the real frequencies, $\omega$, and summing over fermionic Matsubara frequencies we arrive at:
\begin{equation}
    \chi_{+-}(\q,\omega)=\sum_{l,l'}(\chi_{RPA}(\q,\omega))^{l\up,l\down}_{l'\up,l'\down},
\end{equation}
in the multiorbital random phase approximation (RPA), where the RPA susceptibility tensor reads:
\begin{equation}
    (\chi_{RPA}(\q,\omega))^{L_1,L_2}_{L_3,L_4}=[\chi^0(\q,\omega)(1-\mathcal{U}\chi^0(\q,\omega))^{-1}]^{L_1,L_2}_{L_3,L_4}.
\end{equation}
Here, $L_i=(l_i,\s_i)$, $\mathcal{U}$ is the Coulomb interaction matrix, and $\chi^0$ the bare susceptibility tensor (\ref{eq:chi0})
provided in Appendix G, now with the $s,t,p,q$ indices running over orbital and spin states. 

\begin{figure}
\includegraphics[width=\columnwidth]{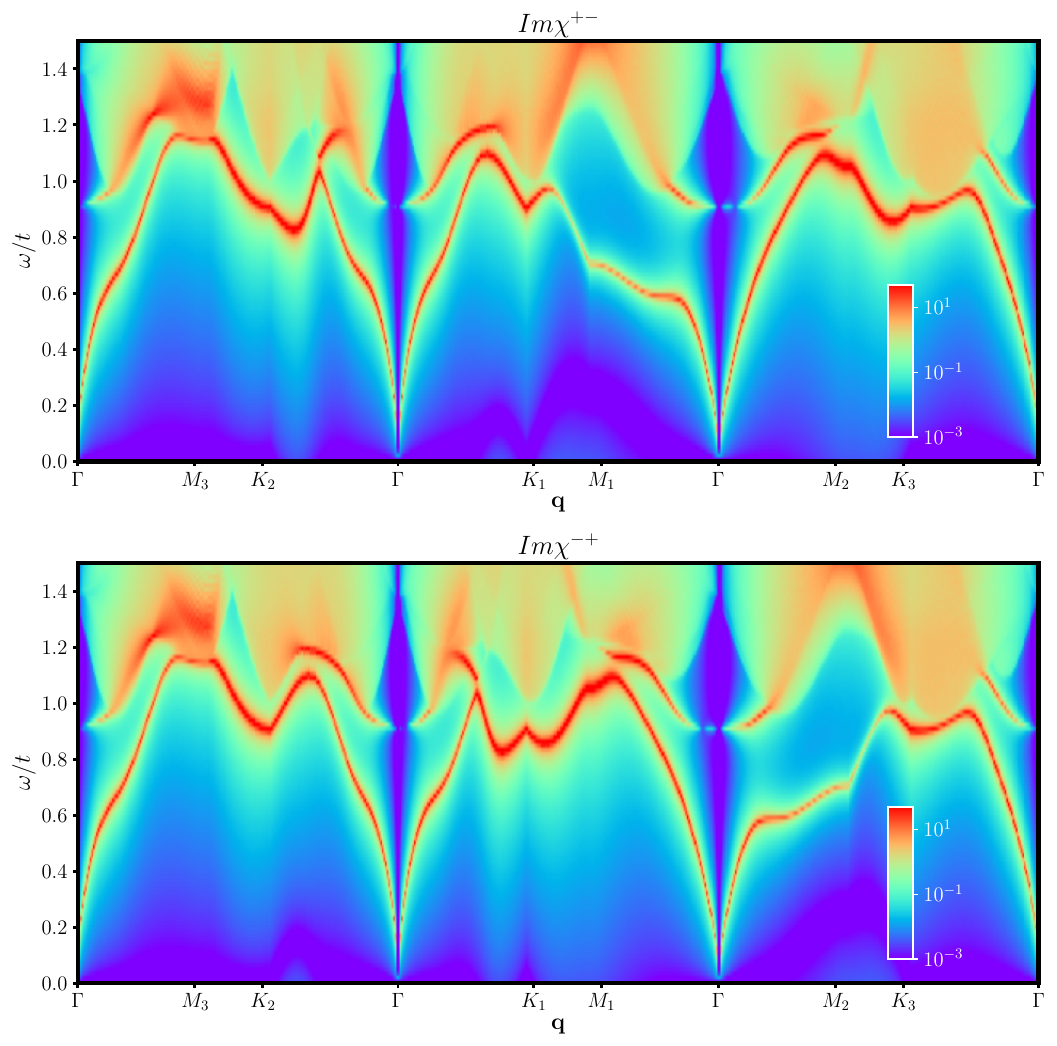}
\caption{\label{chiij} Imaginary part of the physical spin susceptibilities $\chi_{+-}(\q,\omega)$ (top) and $\chi_{-+}(\q,\omega)$ (bottom) of the Kagome altermagnet ($U=8t$, $t'=0.7t$, $n=2/3$, $T=0.01t$) along a momentum path connecting high-symmetry points.}
\end{figure}

In Figure \ref{chiij} we show the resulting spin susceptibilities of the Kagome altermagnet. The INS signal probes the creation of $S=1$ and $S=-1$ magnons which, due to the spin-splitting of the ALM bands, will not have the same energy. Indeed, we find that $\chi_{+-}\neq\chi_{-+}$ along $\Gamma\rightarrow K_{1,2}$, but $\chi_{+-}$ along $\Gamma\rightarrow K_{1(2)}$ is identical to $\chi_{-+}$ along $\Gamma\rightarrow K_{2(1)}$. This is evidenced by the symmetry of the averaged spectrum (main text) with respect to $\Gamma$ and by the splitting of its branches, which is larger along the $\Gamma-M$ than the $\Gamma-K$ paths. On the other hand, there is no magnon splitting 
along the $\Gamma\rightarrow M_3\rightarrow K_2$ or $K_3\rightarrow\Gamma$ paths 
since the ALM band structure is spin-degenerate (e.g., $\Gamma\rightarrow M_3\rightarrow K_2$ or $K_3\rightarrow\Gamma$ paths) implying
$\chi_{+-}=\chi_{-+}$.


\bibliography{bibliography}